\documentclass[twocolumn,preprintnumbers,amsmath,amssymb,pra]{revtex4}
\usepackage{graphicx}
\usepackage{epsfig}
\usepackage{epsf}
\usepackage{amssymb}
\usepackage{dcolumn}
\usepackage{bm}
\usepackage{amsmath}
\usepackage{subfigure}
\usepackage{setspace}
\usepackage{ulem} 
\usepackage{color} 
\usepackage[bookmarks=true]{hyperref}

\newcommand{\half}{\frac{1}{2}}
\newcommand{\mket}[1]{\vert{)1}\rangle}
\newcommand{\mbra}[1]{\langle{)1}\vert}
\newcommand{\ket}[1]{\vert{#1}\rangle}
\newcommand{\bra}[1]{\langle{#1}\vert}
\newcommand{\tr}[1]{\textrm{Tr}\left[{#1}\right]}

\newcommand{\ham}{\mathcal{H}}

\newcommand{\figref}[1]{Fig. (\ref{#1})}

\graphicspath{{../../figures/FAP/}}
\def\ThisFile{\jobname}

\begin{document}
\title {Dynamics and Control of a quasi-1D Spin System}
\author{P. Cappellaro$^1$, C. Ramanathan$^2$ and D. G. Cory$^2$}
\affiliation{$^1$ITAMP - Harvard-Smithsonian Center for Astrophysics, Cambridge, MA 02138, USA\\ $^2$Massachusetts Institute of Technology, Department of Nuclear Science and Engineering, Cambridge, MA 02139, USA\bigskip}
\begin{abstract}
We study experimentally a system comprised of linear chains of spin-1/2 nuclei that provides a test-bed for multi-body dynamics and quantum information processing. This system  is a paradigm for a new class of  quantum information devices that can perform particular tasks even without universal control of the whole quantum system. We investigate the extent of control achievable on the system with current experimental apparatus and methods to gain information on the system state, when full tomography is not possible and in any case highly inefficient.
\end{abstract}
\maketitle

\section{Introduction}
The dynamics and control of complex quantum many-body systems has elicited renewed interest within the field of quantum information science. Not only is there a need to improve coherent control of an increasing number of qubits  to build a scalable quantum computer, but also new task-oriented devices (such as quantum information transport, quantum cloning and simulation devices) constitute a new paradigm for quantum information processing (QIP), in which even a partial control over the system is enough to accomplish the desired task. Of course, even for these devices a precise knowledge of their dynamics is an indispensable ingredient, thus spurring new investigations and the development of new tools for coherent control.

Here we study a physical system that presents a simpler quantum dynamics because of its one-dimensional geometry, while still retaining the full complexity of many-body dynamics. The system is a single crystal of fluorapatite (FAp), where we focus on the spin degrees of freedom of the fluorine-19 nuclei.  This spin system has already been studied in the nuclear magnetic resonance (NMR) literature and proposed for quantum computation. 

The goals of this study are to investigate the extent of our control on the system, when the control Hamiltonians can only act collectively on all the spins; and  to devise methods for extracting information on the system, when state tomography is not only inefficient, but also impossible because we can only detect the collective magnetization. We do not consider here issues of decoherence mechanisms and rates, relying on the long coherence times of nuclear spins for performing all experiments in the coherent regime. Decoherence properties of quantum many-body systems is however a very important area of study \cite{Suter,MQCDecayTheory,LossSpinDecay}, and preliminary results regarding the system under investigation have already been presented \cite{JoonPRB}.
 
 In section II we first show how we can push the limits of our control when preparing a particular initial state, which is of interest for some tasks such as quantum information transfer and quantum simulations \cite{Giovannetti1}. Starting from the thermal equilibrium state, we will show how to create a state where only the two spins at the ends of the chain are polarized, while all the other spins are in the identity state. Also, this state is the first step toward showing universal control on the system, as it has been proved that collective control on a qubit chain plus control of the spins at the chain ends  is enough for universal control of the system \cite{fitzsimonsSpinChain,FitzsimonsJones}. In section III we present the characteristics of the physical system and the experimental apparatus as well as the first indications that we successfully prepared the desired state. 
 We then explain  in section IV how we can extract enough information from the detection of the collective magnetization, to verify the creation of this desired initial state. The measurement is preceded by evolution under a propagator engineered to let emerge the many-body properties of the system. We use the technique of multiple quantum coherence (MQC)  \cite{MQC} as a way to perform a partial tomography of the system and gain further information on its state. In particular, we take advantage of the lower dimensionality of the system, which makes possible an analytical description of the evolution. 
We finally present  experimental results in good agreement with the theoretical predictions, thus proving the preparation of the desired state. 

\section{Selecting the spins at the extremities of the linear chain} \label{SelEndChain} 
There has been recently a remarkable interest in linear chains of spins \cite{burgarthbose,paternostro,Bose,Kay,Giovannetti}, as well as their simulation through atomic lattices \cite{atomicToolbox}. The physical system we are interested in (a crystal of fluorapatite) can be modeled by a linear chain of spins-1/2 particles of a size ($N\gtrsim10$) such that many-body properties emerge. The system is put at room temperature in a large magnetic field and is addressed through radio-frequency (rf) pulses. The internal Hamiltonian of the system in a frame rotating at the larmor frequency is given by the secular part of the dipolar Hamiltonian:
\begin{equation}	
\label{DipHam}
 	\ham_\textsf{dip}=\sum_{ij}b_{ij}[\sigma_z^i\sigma_z^j-\frac{1}{2}(\sigma_x^i\sigma_x^j + \sigma_y^i\sigma_y^j)]	,
\end{equation}
while external control is provided by the rf Hamiltonian:
\begin{equation}	
	\label{RfHam}	\ham_\textsf{rf}=\omega_\textsf{rf}(t)\sum_{i}[\sigma_x^i\cos{\phi(t)}+\sigma_y^i\sin{\phi(t)}]
\end{equation}
The control is therefore collective, that is, we can only address all the spins in the chain together: This limitation precludes universal control on the system. A recent scheme for QIP \cite{fitzsimonsSpinChain,FitzsimonsJones} showed however that adding to the collective control the ability to manipulate the spins at the chain ends enables universal quantum computation. 
A step toward achieving universal control would be to demonstrate at least a partial addressability of an individual spin. To prove the extent of our control on the system, we set up to prepare a particular quantum state, in which only the spins at the extremities  of the chain are polarized: $\rho_0=\openone+\epsilon(\sigma_z^1+\sigma_z^N)$ (notice that this highly mixed state emerges from an ensemble average over many chains that are here considered as independent and equivalent systems). This state not only proves that we can break the internal symmetry of the system, but is also a useful state, for example for quantum information transport \cite{TransportPaola}.

The spins at the extremities of the chain only have one nearest neighbor to which they are strongly coupled. This implies  a different dynamics under the internal dipolar Hamiltonian with respect to the other spins. 
Taking advantage of this fact and complementing unitary control with  incoherent control  \cite{Dalessandro} will allow us to reach the desired state with a high enough fidelity. 

The spin system is initially at equilibrium, in the thermal state; in the high field, high temperature limit this can be very well approximated by the state $\rho_{eq}=\sum_{k=1}^N \sigma_z^k$ (notice that we ignore the large component proportional to the identity since it does not evolve nor contribute to the signal).
By rotating the thermal equilibrium state to the transverse plane $\rho(t=0^+)=\sum_{k=1}^N \sigma_x^k$, the state is no longer an eigenstate of the internal Hamiltonian and will therefore evolve. This evolution is  usually recorded in the NMR  free induction decay (FID). It is known that the apparent decay of the magnetization masks a complex dynamics \cite{JoonFid}, where non-detectable many-body states are created from the single-body magnetization terms by the  dipolar Hamiltonian couplings. Due to the fewer number of couplings to other spins, the first and last spins have a much slower dynamics (apparent decay) at short time. It is thus possible to select a particular time $t_1$ at which while the state of these two spins is still mainly $\sigma_x$, all the other spins have evolved to more complex multi-body states. 
From the commutator expansions of the unitary evolution, we can calculate the approximate coefficients of the polarization ($\sigma_x^k$) terms for each spin as a function of time and therefore select the time at which $\sigma_x^k\approx0$, $\forall\ k\neq1,N$ (because we are interested in the short time regime, an expansion to the $8^{th}$ order gives an excellent  approximation). 
Except for very short chains (3-4 spins),  the polarization of all the spins $k>2$ is almost equal after this very short evolution and the optimal time  is nearly independent of the number of spins in the chain, therefore allowing us to choose the time $t_1$ even without knowing the precise (average) number of spins in a chain. 

A second $\pi/2$ pulse will bring the magnetization of spins $1$ and $N$ to the longitudinal ($z$-)axis, so that the density matrix describing the system can be written as $\rho(t_1)=\alpha (\sigma_z^1+\sigma_z^N) + \rho'$. To select only the first two terms, which are the desired state, we can recur to a phase cycling scheme that selects only terms that commute with the total magnetization along $z$ ($\sum_{k=1}^N \sigma_z^k$) such as population terms. Unfortunately, we have not found a solution that also cancels out the zero-quantum terms (that is, components of the density matrix with total magnetic quantum number=0). Even with this limitation, the fidelity with the desired state is about 70\%; the larger errors are given by residual polarization on spins 2 and N-1 as well as correlated states of the form $\sigma_z^i(\sigma_+^{i-1}\sigma_-^{i+1}+\sigma_-^{i-1}\sigma_+^{i+1})$.

\section{Experimental setup}
We performed  experiments on a 300MHz Bruker Avance Spectrometer, with a home built probe tuned to 282.4MHz for the observation of fluorine spins. The sample was a single crystal of fluorapatite (Ca$_5$(PO$_4$)$_3$F). 
Apatites, either  hydroxyapatites or fluorine containing apatites \cite{FAP2exp,FAP1exp}, have been studied in NMR experiments because of their particular geometry \cite{FAPYesi1,FAPYesi2} and have also been proposed as a system to implement QIP \cite{FAPQIP}. 
The fluorapatite crystallizes in the hexagonal-dipyramidal crystal system,
 with cell dimensions $a = 9.367\textrm{\AA}$ 
and c=$6.884\textrm{\AA}$, and two formula units per cell. 
The fluorine spins are arranged on linear chains along the $c$ direction, with distance between two atoms 
$d=3.442\textrm{\AA}$, and 6 adjacent parallel chains at the distance $D=9.367\textrm{\AA}$. Since the dipolar couplings decrease with the cube of the distance between spins, the spin system can be considered a quasi-1D system. In particular, by orienting the $c$ axis of the crystal along the  $z$ direction, the ratio of the cross-chain to in-chain dipolar coupling is:
\begin{equation}
	\frac{b_\times}{b_{in}}=\half \frac{d^3}{D^3}\approx 0.0248	.
\end{equation}

Although this crystal has a very good purity, as testified by long decoherence times ($T_1\sim200s$), the chains are interrupted by defects and vacancies, that limit their length.
 If we consider the short-time evolution only, the spins of different chains do not interact and behave as independent systems, while if we let the system evolve for a time long  compared to the cross-chain interactions, the approximation of independent systems will fail.
We will include the effects of other chains and the distribution in chain length into the environmental decoherence  and only consider a single chain as the system of interest . 
 
We applied the pulse sequence presented in the previous section:
\begin{equation}\label{pulseSeq}
\left.\frac{\pi}{2}\right|_{\alpha} - t_1 - \left.\frac{\pi}{2} \right|_{\bar{\alpha}},
\end{equation}
where the pulse axis $\alpha$ was phase cycled through $y$ and $x$ to cancel the non zero-quantum terms.  The optimal time $t_1$ for the dipolar coupling strength of fluorapatite spin chains is given by $t_1=30.3\mu s$. 

In order to assess qualitatively the results of this sequence, we measured a spectrum of the system, after the preparation of the desired initial state and compared it to the thermal equilibrium spectrum.
 In particular, it is important to observe the state at short time, when the effects of the desired initial state should be stronger; the solid echo technique \cite{SolidEcho} was thus used, to avoid the dead time imposed by the electronics and the pulse ring-down \cite{Fukushima}. While this kind of measurement does not give a definite answer to the question whether the polarization has been concentrated on the extremities only, the qualitative differences in the spectra measured are encouraging. 
In particular, we observe that resolution of the three peaks observed in the thermal state is much better 
\begin{figure}[bth]
	\centering
		\includegraphics[scale=0.3]{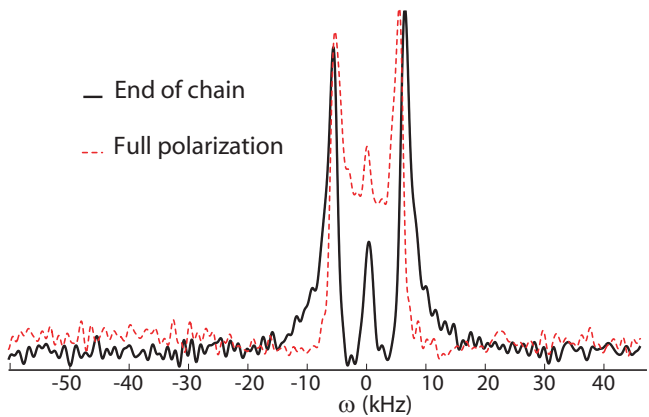}
	\caption{\textbf{Comparison of the spectra} when the polarization is retained by all spins in the chain (dashed line) and for the excitation of the extremities only (solid line). The FWHM is $\approx 19kHz$ and the distance between peaks $\approx 8kHz$. The experimental data were obtained with a modified  version of sequence \eqref{pulseSeq} and a Solid Echo read-out. (The sequence was equivalent to the one presented, but  an extra pulse gave experimentally a cleaner spectrum when combined to the read-out sequence). The wait time $t_1$ was $t_1=.5	\mu s$ for the all spin spectrum and $30.3\mu s$ for the chain ends excitation.}	\label{SelEndCh_Spectra}
\end{figure}
(we obtain  narrower lines) as expected from a state in which fewer couplings are available, and while the spectrum of the thermal state presents the characteristic of a triplet \cite{FAP1exp,FAP2exp} as expected for one spin coupling strongly to two equivalent spins (as is the case if we consider nearest neighbors only) the spectrum for our state is closer to a doublet, reflecting the fact that the spins at the chain extremities interact strongly with only one spin.

\section{Multiple Quantum Coherence as a state measurement tool.} \label{Verifying the state preparation}  
Since the detection of magnetic resonance  restricts the observables to the collective transverse magnetization, it is not possible to reconstruct the state of the system by quantum state tomography \cite{tomography}. To assess the efficiency of the state preparation scheme, we thus need a read-out scheme that reveals the signature of the particular initial state prepared.
By studying a more complex dynamics than the free induction decay, we can obtain more accurate insight into the state created. In particular, a sensitive probe of the dynamics of a correlated many-spin system is  the creation and evolution of  quantum coherences. 

\subsection{Fermion operator solution to MQC dynamics} \label{MQC dynamics for the two distinct initial states}  

Quantum coherence refers to a state of a physical system where the phase differences among the various constituents of the  wave function can lead to interferences. In particular, quantum coherences often refer to a many-body system, whose parties interact and therefore show a correlation, a well defined phase relationship. 
In NMR, coherences between two or more spins are usually called \textit{multiple quantum coherences}.
When the system is quantized along the z axis, so that the Zeeman magnetic moment along z is a good quantum number, a quantum coherence of order $n$ is defined as the transition between two states $\ket{m_1}$ and $\ket{m_2}$, such that the difference of the magnetic moment along z of these states is $n$: $m_1-m_2\propto n$. Multiple quantum coherences of order $n$  usually describe states like $\ket{m_2}\bra{m_1}$, or elements in the density matrix that correspond to a transition between these two states \cite{vega,MQCbook}. The state $\ket{m_2}\bra{m_1}$ is also called a coherence of order $n$. Diagonal density matrix terms are called populations and do not properly describe a coherence.

The observation of multiple quantum coherences in NMR started in the mid 1970s, as a method for unraveling  complex spectra, by filtering transitions based on the coherence order involved \cite{MQCbook}.  More recently, these states have been studied with respect to their decay time  and effects of decoherence on them \cite{JoonPRB,Suter}, because of their connection to entangled states. Multiple quantum coherences are also the basis of proposed schemes for the full characterization of complex many-spin states \cite{MQCSpherical}: identifying the different coherence orders occurring in a state is the first step toward quantum state tomography.

Coherences can be created by the interplay of rf pulses and free evolution periods, during which interactions among spins occur. Multiple pulses sequences create Hamiltonians that can raise the coherence order of the system. One widely used sequence \cite{MQC,MQCxz}, for example, creates the double-quantum Hamiltonian
\begin{equation}
\label{DQHamiltonian}
 \mathcal{H}_{DQ}=\sum_{i,j}\frac {b_{ij}}{2} (\sigma^x_i\sigma^x_j - \sigma^y_i\sigma^y_j) = \sum_{i,j}b_{ij}(\sigma^+_i\sigma^+_j + \sigma^-_i\sigma^-_j)
 \end{equation}
which raises the coherence orders in step of two (thus, starting from the thermal state, only even coherence order will be created). The growth of coherence orders has been the matter of many investigations \cite{SpinCounting2,JoonFid}, because of its relationship to the system's geometrical structure. Because of its complexity, only stochastic models, based on the probability of occupation of different coherent states, and semiclassical models like the hopping model are available to describe this dynamics. If we restrict the evolution to one-dimensional systems, the growth of coherence is slowed down by the fewer couplings among spins; if we further assume that only nearest-neighbor couplings are present, the evolution under the double-quantum Hamiltonian turns out to be exactly solvable \cite{Lacelle, Feldman}. Experimentally, the nearest-neighbor (nn) approximation is accurate for short times, while for longer times, weaker couplings start to produce appreciable corrections. 

The most important characteristic of 1-D MQC experiments are the oscillations between zero- and double-quantum coherences at short times. It is this restriction of the accessible Hilbert space (that is exact at any time in the nn approximation) that makes the problem analytically tractable. As we will show in this section, these oscillations turn out to be a signature of the initial state, so that they can help us confirm experimentally the preparation of the initial state desired.
For comparison, we derive analytically (and then measure experimentally) not only the MQC intensities for the desired state  $\rho_e(0)=\sigma^1_z+\sigma^N_z$, but also for the thermal equilibrium state $\rho(0)=\sum_{j=1}^N \sigma^j_z$. 

The analytical result is obtained by mapping the spin system to spinless fermion operators (see the appendix):
\begin{equation}
	\label{fermionicsimple}
	c_j=-\prod_{k=1}^{j-1}\left(\sigma^k_z\right) \sigma_j^-
\end{equation}
 The diagonalization of the double-quantum Hamiltonian \cite{Lieb,Lacelle,Feldman} is accomplished  by a Bogoliubov canonical transformation to the operators $d_k$ \cite{bogoliubov}:
\begin{equation}
	\label{FermBog1}
	c_j=\displaystyle\frac{1}{\sqrt{N+1}}\sum_{k=1}^N\sin{(\kappa j)}~(\gamma_k d_k+d^\dag_{-k}),\ \ \kappa=\frac{\pi k}{N+1}
\end{equation}
where $\gamma_k\equiv\textrm{sgn}(k)$. 
The $DQ$-Hamiltonian is then diagonalized to:
\begin{equation}
	\label{DQHamBog}
	\ham_{DQ}=-2b \sum_{k=1}^N\cos{\kappa}~(d^\dag_kd_k+d^\dag_{-k}d_{-k}-1)
\end{equation}
 
We now express the two initial states  in terms of Bogoliubov operators $d_k$. 
The thermal state has a particularly compact expression.
Using first the spin to fermion mapping, we have
\begin{subequations}
\label{RhoFermFour}
\begin{gather}
\rho(0)=\half\sum_{j=1}^N\sigma_j^z=
\sum_{j=1}^N \left(\half-c^\dag_jc_j\right)\label{RhoFermFour1}\\
=\frac{N}{2}\sum_{k=1}^N (d_kd_{-k}-d^\dag_kd^\dag_{-k})\label{RhoFermFour2}
\end{gather}
\end{subequations}
where in writing \eqref{RhoFermFour2} we used orthogonality relationships for trigonometric functions (see Eq. \eqref{trig1} and \eqref{trig2} in the appendix).

Consider now the initial state $\rho_e(0)=\half(\sigma_1^z+\sigma_N^z)$. Because we are not summing over all spins, it is no longer possible to use the orthogonality relationships as in \eqref{RhoFermFour2}. This results in more cumbersome double sums: 
\begin{subequations}
\label{RhoEndFermFour}
\begin{gather}
\rho_e(0)=\half(\sigma_1^z+\sigma_N^z)=1-(c^\dag_1c_1+c^\dag_Nc_N)\\
=1-
\displaystyle\frac{1}{N+1}\sum_{k,h}S_{kh}(\gamma_kd^\dag_k+d_{-k})(\gamma_hd_h+d^\dag_{-h}),
\end{gather}
\end{subequations}
where to simplify the notation we set $S_{kh}=\sin(\kappa)\sin(\eta) + \sin{(N\kappa)}\sin{(N\eta)}$. 

The system evolves under the double quantum Hamiltonian with a dynamics described by the propagator $U(t)=\exp{(2ib\sum_k \cos{\kappa}(d^\dag_kd_k+d^\dag_{-k}d_{-k}))}$. 
We now define the eigenphases $\psi_k=2bt\cos{\kappa}$. 
 The thermal state evolution is easily calculated to be:
 \begin{equation}
 \label{RhoEvol}
\rho(t)=\half\displaystyle\sum_{k=1}^N\left(d_kd_{-k}e^{2i\psi_k}-d^\dag_kd^\dag_{-k}e^{-2i\psi_k}\right)
 \end{equation}
 In order to separate contributions from different coherence orders, we transform back to  fermion operators. To simplify the calculations, we use the operators  $a_k=(\gamma_kd_k+d^\dag_{-k})/\sqrt{2}$ that represent the same coherence order as the $c_j$'s operators (the  $d_k$ operators correspond instead to different coherence orders, since they are  combinations of lowering and raising operators). With some algebraic manipulations, we have
 \begin{equation}\label{RhoEvol2}
 \begin{array}{ll}
 \rho(t)=&\underbrace{-\sum_k\cos{\psi_k}(a_k^\dag a_k-\half)}_{\rho^{(0)}}\\&-\underbrace{\frac{i}{2}\sum_k\gamma_k\sin{|\psi_k|}(a_k^\dag a^\dag_{-k}+a_k a_{-k})}_{\rho^{(+2)}+\rho^{(-2)}}
 \end{array}\end{equation}
 The intensities $J_n$ of each $n^{th}$ quantum coherence as measured in MQC experiment is given by $\tr{(U_{DQ}\rho(0)U_{DQ}^\dag)^{(n)} (U_{DQ}\sum \sigma_z U_{DQ}^\dag)^{(n)}}$. Since in this case $\rho(0)=\sum\sigma_z$, we have  $J_n=\tr{\rho^{(n)}\rho{^{(-n)}}}$. We evaluate the trace of the fermion operators $a_k$, $a_k^\dag$ in their corresponding occupational number representation, so that only terms like $a^\dag_ka_k$ are diagonal and contribute to the trace (see Eq. \eqref{TraceFermFour1} and \eqref{TraceFermFour2} in the appendix).  The normalized MQC intensities for zero and double quantum are finally given by \footnote{Notice the discrepancy with the result in \cite{Lacelle} which is due to incorrect boundary conditions. We confirmed our results with numerical calculations for short chains.}:
 \begin{subequations}\label{MQCIntTh}
 \begin{gather}
 J_0=\frac{1}{N}\sum_k \cos^2{(4bt\cos k)} \\
 J_2=\frac{1}{2N} \sum_k \sin^2{(4bt\cos k)}
 \end{gather}\end{subequations}

The evolution of the polarization on the extremities of the chain is :
\begin{equation}
	\label{EndEvolBog}
	\begin{array}{l}
	\rho_e(t)=1-\displaystyle\frac{1}{N+1}\sum_{k,h=1}^NS_{kh}
\\	\times\left( d^\dag_k e^{-i\psi_k}+ d_{-k}e^{i\psi_k}\right)\left( d_he^{i\psi_h}+d^\dag_{-h}e^{-i\psi_h}\right)\end{array}
\end{equation}

Again, we transform to fermion operators to distinguish contributions from different coherence orders:
\begin{equation}
	\label{EndEvolBog0}
	\begin{array}{l}
\rho_e(t)
=\underbrace{ \frac{i}{2}\frac{1}{N+1}\sum_{k,h}S_{kh}\sin{(\psi_k+\psi_h)}(a^\dag_ha^\dag_{-k}-a_{-k}a_h)}_{\rho^{(+2)}+\rho^{(-2)}}\\
\underbrace{+1-\frac{1}{N+1}\sum_{k,h}S_{kh}[a^\dag_ka_h~\cos{(\psi_k+\psi_h)}+\delta_{k,h}\sin{\psi_k}^2]}_{\rho^{(0)}}
\end{array}
\end{equation}

As expected, also starting from the non collective initial state only zero and double quantum coherences are developed (taking into account nearest neighbor couplings only). 

The zero quantum contribution to the signal is $J_0^\textsf{e}=\tr{\rho_e^{(0)}\rho^{(0)}}$ (with $\rho^{(0)}(t)$ from eq. \eqref{RhoEvol2}):

\begin{equation}
\label{ZQEndFinal1}
\begin{array}{l}
J_0^\textsf{e}=-\sum_k\cos{\psi_k}\tr{a_k^\dag a_k-\half}\\
+\frac{1}{(N+1)}\displaystyle\sum_{k,h,l}S_{kh}\cos{\psi_l}\cos{(\psi_k+\psi_h)}\tr{a^\dag_ka_ha^\dag_la_l}\\
+\frac{1}{(N+1)}\displaystyle\sum_{k,l}S_{kk}\cos{\psi_l}\sin{\psi_k}^2~\tr{a_k^\dag a_k-\half}
\end{array}
\end{equation}
Using the trace and  trigonometric identities in the appendix, we find the normalized zero-quantum intensity:
\begin{equation}\label{MQCIntEnd0}
J_0^\textsf{e}=\frac{2}{(N+1)}\sum_{k}\sin^2{(\kappa)}\cos^2{(4bt\cos{\kappa})}
\end{equation}%

\begin{figure}[htb]
	\centering
		\includegraphics[scale=.25]{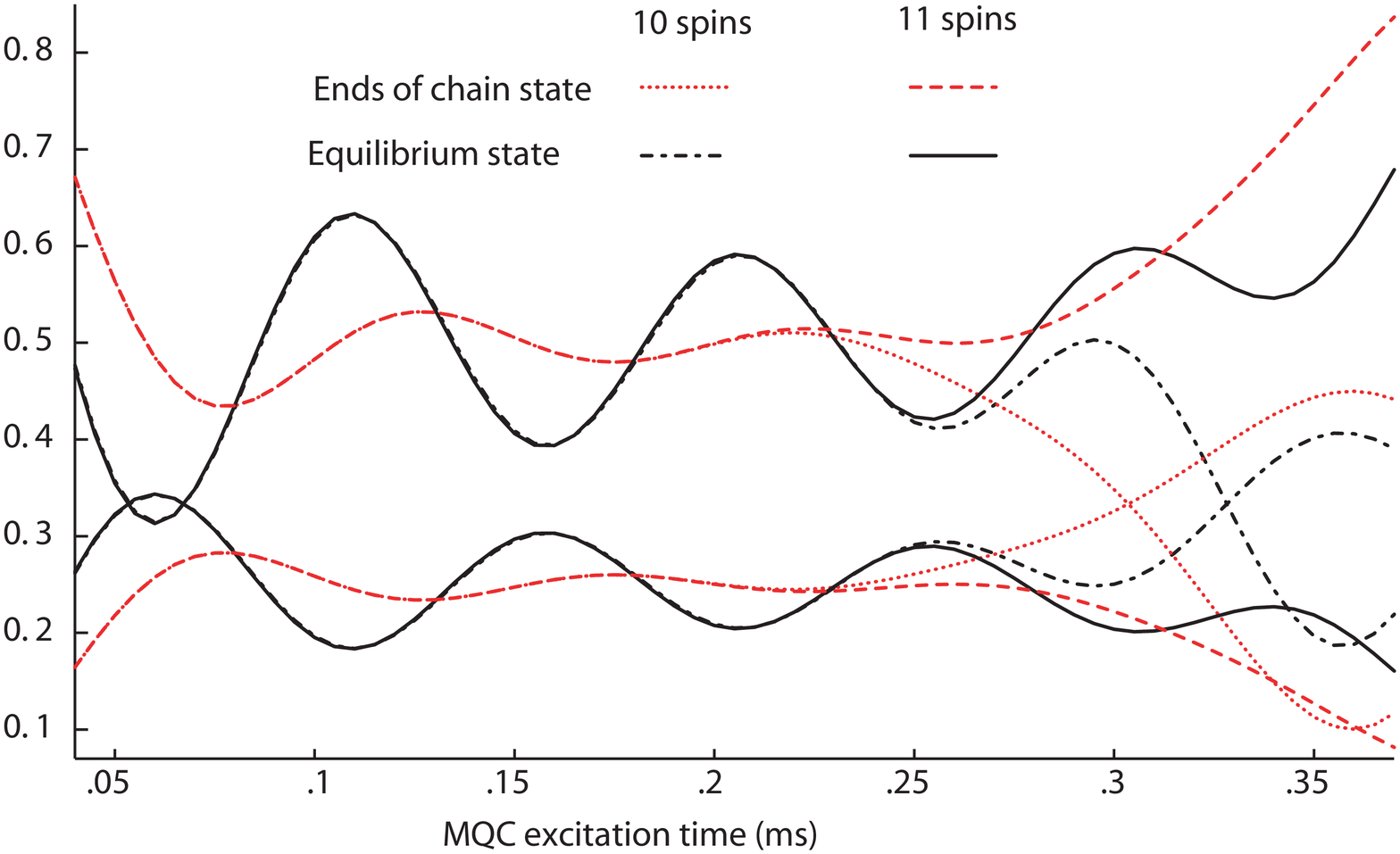}
	\caption{\textbf{Zero- and double-quantum intensities} as a function of the evolution time under the double-quantum Hamiltonian. Nearest-neighbor couplings only are assumed, with equal strength as given by the fitting to experimental data (see \figref{MQCExpSim}). In particular notice the clear differences in the behavior for the two initial states. Also the even-odd spin number dependence of the MQC intensities is interesting: while this tends to go to zero for large number of spins in the collective initial state case, this difference is observed even for very large number of spins for the other initial state. }
	\label{MQCExpThy}
\end{figure}

The double quantum intensity $J_2^\textsf{e}$ is given  by 
\begin{equation}
\label{DQEnd1}	
\begin{array}{lr}
\displaystyle J_2^\textsf{e} &=\tr{\rho_e^{(2)}\rho^{(-2)}}= \frac{1}{(N+1)}\sum_{k,h,l} S_{kh} \sin{\psi_l}\\&
\times\half\sin{(\psi_k+\psi_h)}\tr{a^\dag_ha^\dag_{-k}a_{l}a_{-l}},
	\end{array}
\end{equation}
which yields:
\begin{equation}
\label{MQCIntEnd2}
\begin{array}{l}
J_2^\textsf{e}=\frac{1}{(N+1)}\sum_{k}\sin^2{(\kappa)}\sin^2{(4bt\cos{\kappa})}
\end{array}
\end{equation}

These more complex expressions lead to a very different behavior of the coherence intensities as shown in \figref{MQCExpThy}, so that it is possible to distinguish even experimentally what was the initial condition of the system.

\subsection{Experimental results}
We applied the pulse sequence \eqref{pulseSeq} followed by a MQC-experiment sequence. In particular, we used the 16-pulse sequence \cite{MQCxz} (except for the 3 shorter time-values, where the 8-pulse sequence was used \cite{MQC}), and a phase cycling with increments of $\varphi=2\pi/8$ to select up to the $4^{th}$ coherence order, by repeating the experiment 16 times. 
\begin{figure*}[hbt]
	\centering
		\includegraphics[scale=.5]{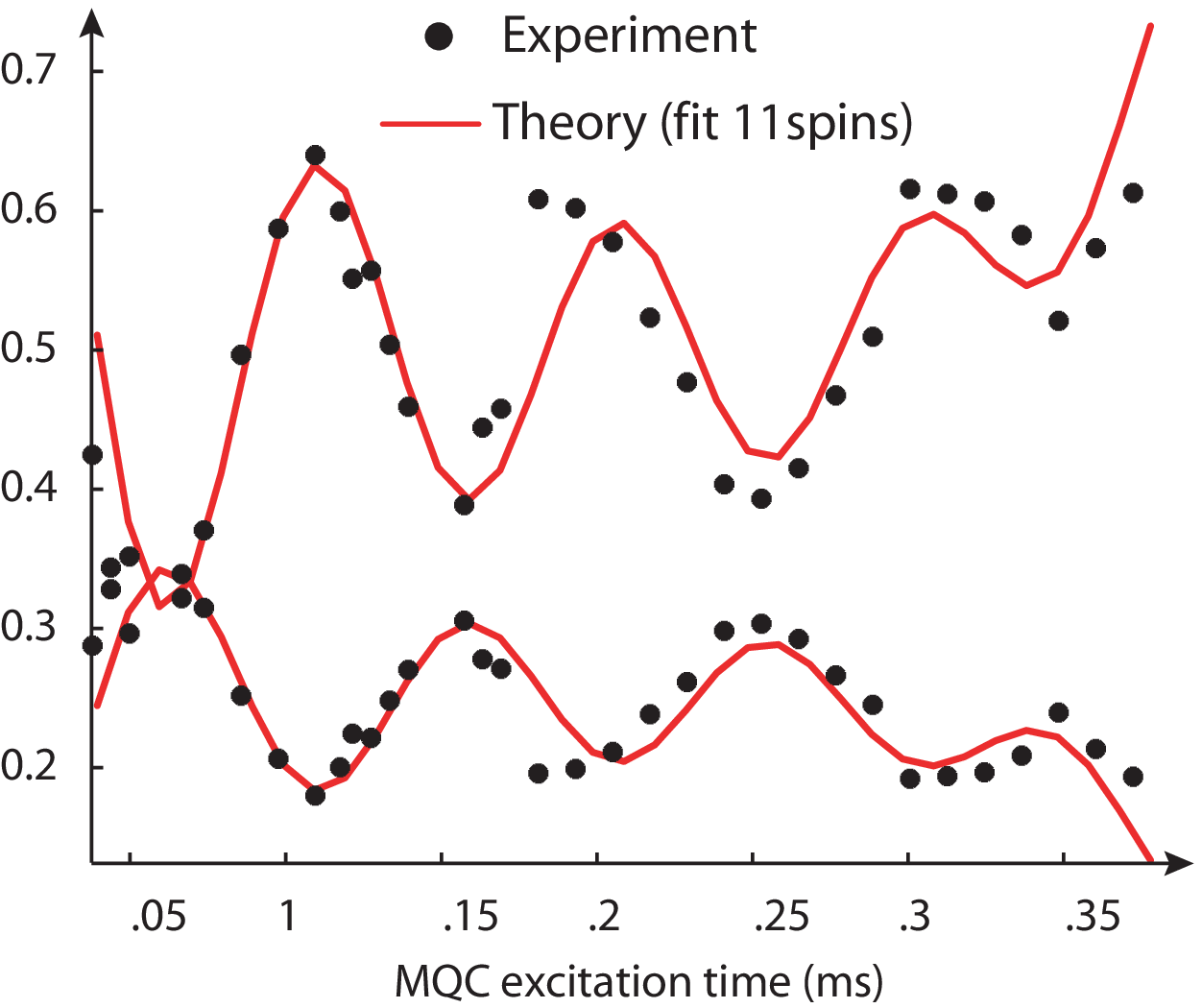}
\hspace{0.1\textwidth}		
		\includegraphics[scale=.5]{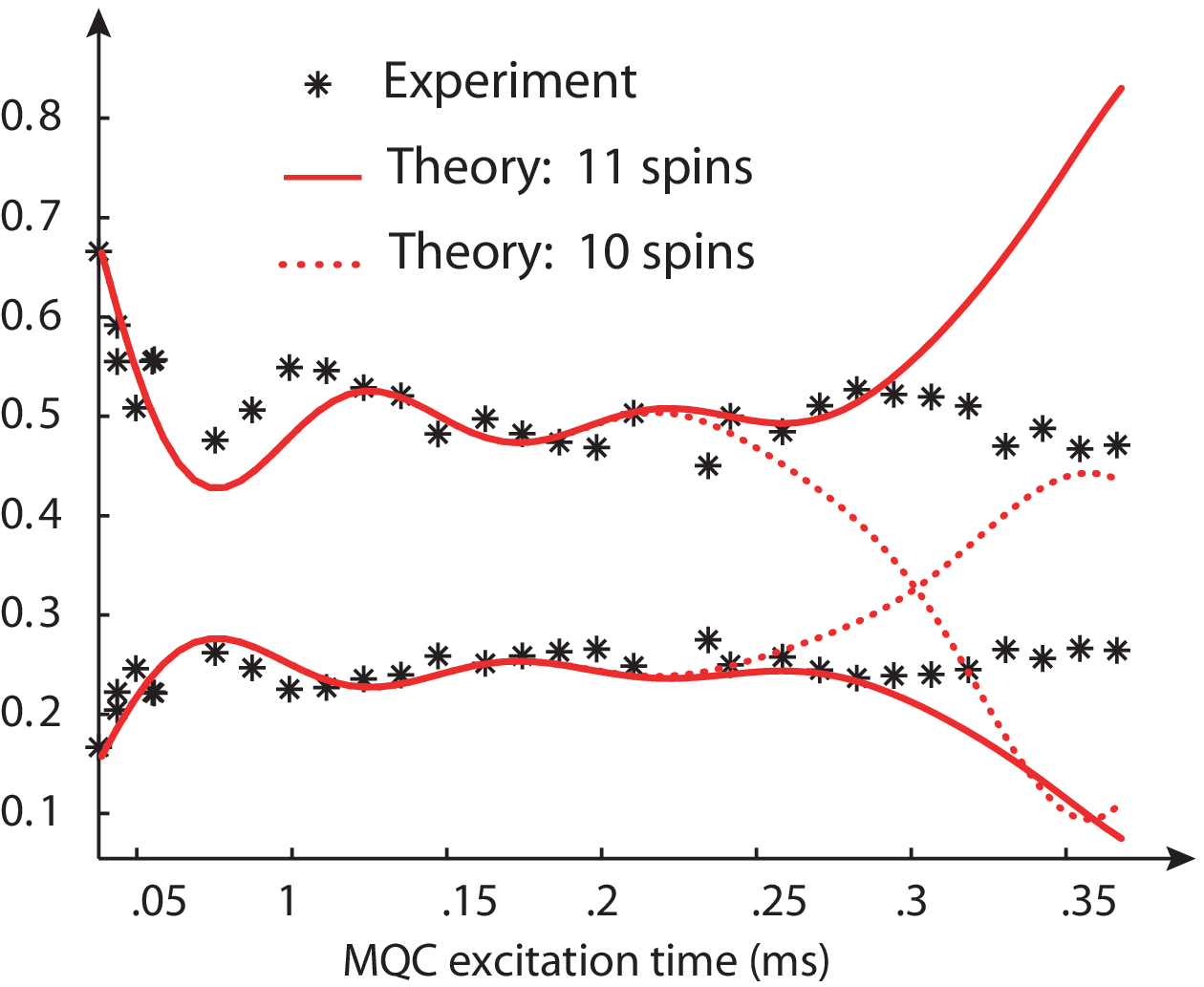}
	\caption{\textbf{Experimental results.} Left: the initial state is the collective thermal state $\sum_k\sigma_z^k$. The experimental points have been fitted (dashed line) to the theoretical curves for nearest neighbor coupling only, with the dipolar coupling as fitting parameter. The number of spins was varied to find the best fit, which results to be $N=11$ spins. Right: MQC intensities for the initial state $\rho_0=\sigma_z^1+\sigma_z^N$. Also plotted are the theory predictions for the same dipolar coupling and 11 spins (solid line) or 10 spins (dashed line). The experimental behavior observed (a constant behavior also for longer time) indicates that a simple model with 11 spin chains is not an accurate description at longer times (see text).}
	\label{MQCExpSim}
\end{figure*}

 The time delay between pulses in the MQC sequence was varied from $2 \mu s$ to $6.5 \mu s$, to increase the excitation time, as well as the number of repetitions of the sequence itself (1 or 2 loops), so that the evolution of the quantum coherences were studied between the times of $37.6 \mu s$ to $354.4 \mu s$. We compare the results obtained with the evolution for the thermal equilibrium as  initial state. In order to take into account the effects of imperfections in the pulse sequence, we applied the pulse sequence as in \eqref{pulseSeq} also to obtain the thermal state, with a very short $t_1$ time ($t_1=0.5\mu s$). In figure \eqref{MQCExpSim} we show the dynamics of the zero and double quantum intensities, normalized to have sum $=1$ to take into account the signal decay for longer excitation times. 
We notice that the four-quantum coherence intensity is as low as the baseline, indicating that the time scale is short enough for the nearest-neighbor approximation to be valid to a good extent (remember that the nn-approximation predicts that only zero- and double-quantum coherences are excited). 

The experimental results for the MQC oscillations starting from the thermal state have been fitted to the theoretical curve  \eqref{MQCIntTh} for a single chain with nearest neighbor couplings only. The dipolar strength and the number of spin were the fitting parameters. The results of this fitting were used to plot the theoretical curve for the chain ends initial state [Eqn. \eqref{MQCIntEnd0} and \eqref{MQCIntEnd2}].
The concordance of the theoretical predictions with the experimental data is very good, even if the state created contains residual zero-quantum terms. 

The best fitting of the experimental data in Fig. 3 to \eqref{MQCIntTh} was found for a chain length of 11 spins. This result must be taken with caution, since the chain length influences only the long time behavior of the MQC intensities, where other factors not taken into account in the simple analytical model start to play an important role. In particular, the experimental results for the chain ends indicate that there are either longer chains or a distribution of short chains, with odd and even number of spins.
Although the reason for the best fit at 11 spins could be simply due to a distribution of spin chains around a mean of 11 spins, the low impurities content of the crystal studied is not in agreement with this finding. We believe that a more plausible reason is the break down of the model of isolated nearest-neighbor coupled spin chains. Even if next-nearest neighbor couplings and couplings to adjacent chains are not strong enough yet to create four-quantum coherences, they still modify the intensities of the zero- and double-quantum coherences. Further experiments would be needed to distinguish between these hypotheses.

\section{Conclusions}
We have studied a naturally occurring physical system, a single crystal of FAp that presents linear chains of spins-1/2 particles.   Since the physical characteristics of the system and the experimental apparatus do not provide universal control on the quantum spin system, we propose to use this system not as a candidate quantum computer, but as a specific task-oriented QIP device, for example for quantum state transport or simulations. In particular, we have devised a scheme, combining unitary and non-unitary control, for creating a particular state that breaks the natural symmetry of the system. This  state will allow us to study properties of state transport along the chain as well as being an interesting initial state for simulations. Furthermore, the preparation of this state is the first step toward universal control on the system, since  full control on the spins at the chain ends (in addition to the collective control over all other spins) ensures  universality. 

In addition, we have investigated a tool for acquiring a deeper knowledge of the state and dynamics of the system, given the limitations in the read-out procedures. Multiple quantum coherences allow us to gather more information on the multi-body aspects of the system than simple direct observation of the collective polarization. In particular, we used analytical solutions in the limit of nearest-neighbor couplings to interpret the experimental results which confirm the preparation of the desired state. 

In conclusion, we have shown how even a quantum system without universal control can be used to study physical problems of interest in condensed matter theory and quantum information science.

\textbf{Acknowledgments}.  
This work was supported in part by the National Security Agency (NSA) under Army Research Office (ARO) contracts DAAD190310125 and  W911NF-05-1-0459, by DARPA and bythe National Science Foundation under Award 0403809 and through a grant for the Institute for Theoretical Atomic, Molecular and Optical Physics at Harvard University and Smithsonian Astrophysical Observatory.

\section{Appendix: Fermion Operators}\label{Fermionic Operators}
Spin operators of the Pauli group can be mapped to fermion operators, obeying the well-known anticommutation relationships. This mapping is useful in describing the dynamics of various 1D models, since some Hamiltonians can then be diagonalized analytically.
In the following we describe a particular mapping that is suited for describing the creation of MQC.

A mapping from spin to fermion operators goes back to Jordan and Wigner
\cite{JW}, who first transformed quantum spin S = 1/2 operators, which commute at different
lattice sites, into operators obeying a Clifford algebra (fermions). This transformation was used
to map the one-dimensional Ising model into a spinless fermion model, which is
exactly solvable. The Jordan-Wigner transformation has been recently generalized   to the cases of arbitrary spin S \cite{Batista,Anfossi} and to 2D spin systems \cite{Verstraete}.

Given a set of spin-$\half$ operators $\sigma_j^\alpha$, each defined at a lattice site $j$, they obey the commutation relationship:
\begin{equation}
[\sigma_j^\alpha,\sigma_k^\beta]=\delta_{j,k}i\sigma_j^\gamma, 
\end{equation}
where $\{\alpha,\beta,\gamma\}=\{x,y,z\}$ and cyclic permutations of these indexes. 
The raising and lowering operators $\sigma_j^\pm=(\sigma_j^x\pm i\sigma_j^y)/2$ obey mixed commutation and anticommutation relationships, which are not preserved under a unitary transformation. To diagonalize the spin Hamiltonian we must thus use the Jordan-Wigner transformations to map these operators to fermion operators $c_j,\ c_j^\dag$, obeying the canonical anticommutation relationships:
\begin{equation}
\{c_j^\dag,c_k\}=\delta_{j,k},\ \ \ \ \ \{c_j,c_k\}=\{c_j^\dag,c_k^\dag\}=0,
\end{equation}
where we adopted the notation $\{\ ,\ \}$ for anticommutators. A basis for the Hilbert space of these operators is given by the occupation number representation $\ket{n}=\ket{n_1,n_2,...,n_N}$, where $n_j=\{0,1\}$ is the occupation number at site $j$. The state $\ket{n}$ can be obtained from the vacuum state by:
\begin{equation}
\ket{n} \equiv \prod_j (c_j^\dag)^{n_j}\ket{vac}.
\end{equation}
Then, the action of the fermion operators on such states is given by:
\begin{equation}
c_j\ket{n}=\left\{ \begin{array}{ll}
0, & \textrm{if }n_j=0\\  -(-1)^{s^n_j}\ket{n'},& \textrm{otherwise} \end{array} \right. 
\end{equation}
where $\ket{n'}$ is the vector resulting when the $j^{th}$ entry of $\ket{n}$ is changed to 0 and $s^n_j=\sum_{k=1}^{j-1}n_k$. Analogously:
\begin{equation}
c^\dag_j\ket{n}=\left\{ \begin{array}{ll}
0, & \textrm{if }n_j=1\\ 
-(-1)^{s^n_j}\ket{n'},& \textrm{otherwise} \end{array} \right. 
\end{equation}
where $\ket{n'}$ is the vector resulting when the $j^{th}$ entry of $\ket{n}$ is changed to 1.

The mapping from spin to fermion operators can be expressed in several ways, the most intuitive being based on identifying every basis vector $\ket{n}$ in the occupation number representation basis to the corresponding $\ket{n}$ basis vector in the computational basis for the spin operator Hilbert space. Imposing this one to one correspondence on basis states and taking into account the respective actions of spin and fermion operators on their basis vectors, one obtains the mapping:
\begin{equation}\begin{array}{cc}
 c_j=-\displaystyle\prod_{k=1}^{j-1}\sigma_k^z~ \sigma_j^-,\ \ &
\sigma_j^-=-\displaystyle\prod_{k=1}^{j-1}\left(1-2 c_k^\dag c_k \right)c_j\\
c_j^\dag=-\displaystyle\prod_{k=1}^{j-1}\sigma_k^z~ \sigma_j^+,\ \  
&
\sigma_j^+=-\displaystyle\prod_{k=1}^{j-1}\left(1-2 c_k^\dag c_k \right)c_j^\dag 
\end{array}\end{equation}
Notice also that $\sigma_j^z=1-2 c_j^\dag c_j$.

Consider now the double quantum Hamiltonian with equal couplings restricted to the nearest neighbor spins :
\begin{equation}
\label{DQHam}
\ham_{DQ}=b\sum_{j=1}^N \sigma^+_j\sigma^+_{j+1}+\sigma^-_j\sigma^-_{j+1},
\end{equation}
We can express it in terms of the fermion operators as:
\begin{equation}
\label{DQHamFerm}\begin{array}{ll}
\ham_{DQ}&=b\displaystyle\sum_{j=1}^N\prod_{l=1}^{j}(1-2 c_l^\dag c_l )c_{j+1}^\dag \prod_{l=1}^{j-1}(1-2 c_l^\dag c_l )c_j^\dag +h.c.\\
&\displaystyle=
 -b\sum_{j=1}^N c^\dag_{j+1}c^\dag_j+c_jc_{j+1}=C^\dag\hat{B}C
 \end{array}
\end{equation}
where we introduced the vector $C^\dag=[c^\dag c]$ and the matrix $\hat{B}$:
\begin{equation}
	\label{Bmatrix}
	\hat{B}=\left[ 
	\begin{array}{cc}
	0 & B \\ -B & 0 
	\end{array}
	\right],\ B=-b(\delta_{i,j+1}-\delta_{i,j-1}) 
\end{equation}
Notice that even if the mapping to fermion operators is non-local, this quadratic Hamiltonian is mapped to a local Hamiltonian, when one considers only nearest neighbor couplings. 
The matrix $\hat{B}$ is orthogonal, and can therefore be put into diagonal form, with eigenvalues $\cos{\left(\frac{\pi k}{N+1}\right)}$, $k=1,2,\dots N$ and eigenvectors $D^\dag=[d^\dag d]$. The diagonalization is a canonical unitary (orthogonal) transformation (a Bogoliubov
transformation) to a diagonal basis satisfying the same anti-commutation relationships \cite{Lieb}. To simplify the diagonalization we need to introduce also the negative modes operators $d_k$ and correspondingly the fermion operators $c_j$ (with $j<0$), by extending the sum in (\ref{DQHamFerm}) from $-N$ to $N$.

We can express the fermion operators $c_j$ in terms of the bogoliubov operators $d_k$ by the following orthogonal relationship:

\begin{equation}
	\label{FermBog}
	c_j=\displaystyle\frac{1}{\sqrt{N+1}}\sum_{k=1}^N\sin{(\kappa j)}(\gamma_k d_k+d^\dag_{-k}),\ \ \kappa=\frac{\pi k}{N+1}
\end{equation}
where $\gamma_k=\textrm{sign}(k)$, ensuring the transformation is orthogonal (this transformation is obtained by imposing that the resulting $DQ$-Hamiltonian is diagonal). 
Substituting expression \eqref{FermBog} in equation \eqref{DQHamFerm} we have:
\begin{equation}
\label{DQHamBog1}
\begin{array}{ll}
	\ham_{DQ} &=\displaystyle - b\sum_{k,h=1}^N[(\gamma_k d_k+d^\dag_{-k})(\gamma_h d_h+d^\dag_{-h})+h.c.] \\
	& \displaystyle\times \frac{1}{N+1}\sum_{j=-N}^N \sin{(\kappa j)}\sin{[\eta (j+1)]}
	\end{array}
\end{equation}
(with $\eta=\pi h/(N+1)$). By using the following orthogonality relationships: 
\begin{equation}
\label{trig1}
\frac{1}{N+1}\sum_{j=-N}^N \sin(kj)\sin(hj)= (\delta_{k,h}-\delta_{k,-h})
\end{equation}
and:
\begin{equation}
\label{trig2}
\sum_{j=-N}^{N} \sin(kj)\cos(hj)= 0
\end{equation}
we can simplify the Hamiltonian \eqref{DQHamBog1}:
\begin{equation}
	\label{DQHamBog2}
	\ham_{DQ}=-2b \sum_{k=1}^N\cos{\kappa}~(d^\dag_kd_k+d^\dag_{-k}d_{-k}-1)
\end{equation}
We note that under this diagonal Hamiltonian, the operators $d_k$ evolve as: $d_k(t)=e^{-i\ham_{DQ}t}d_ke^{i\ham_{DQ}t}=e^{-2ib\cos{\kappa}}d_k$. 

For convenience, we introduce also fermion operators $a_k$ that are useful in expressing the density matrix:
\begin{equation}
	\label{FermFour}
	a_k=(\gamma_kd_k+d^\dag_{-k})/\sqrt{2}
\end{equation}

The following identities for the trace of fermion operators have been used in the text to calculate the zero- and double-quantum intensities :
 \begin{equation}
\label{TraceFermFour1}\begin{array}{l}
\tr{a^\dag_ka_k}=2^{N-1}\\
\tr{a^\dag_ha_ha^\dag_{k'}a_{k'}}=\left\{\begin{array}{ll}
2^{N-2} &\textrm{for }k\neq k';\\
2^{N-1} &\textrm{for }k=k'.
\end{array}
\right.
\end{array}
\end{equation}

\begin{equation}
\label{TraceFermFour2}
\tr{a^\dag_ha^\dag_{-k}a_{-k'}a_{h'}}=\left\{\begin{array}{ll}
2^{N-2} &\textrm{for }k=k',h=h';\\
-2^{N-2} &\textrm{for }k=-h',h=-k';\\
0 &\textrm{otherwise}
\end{array}
\right.
\end{equation}

\bibliography{../UpdateBib}

\end{document}